# Superconductivity in graphite intercalation compounds with sodium

Chun-Mei Hao[1,2], Xing Li[2], Artem R. Oganov[3], Jingyu Hou[4], Shicong Ding[2], Yanfeng Ge[2], Lin Wang[1,2], Xiao Dong[4], Hui-Tian Wang[5], Guochun Yang[2,*], Xiang-Feng Zhou[1,2,4*], and Yongjun Tian[1]

[1]*Center for High Pressure Science, State Key Laboratory of Metastable Materials Science and Technology, Yanshan University, Qinhuangdao, 066004, China*

[2]*Key Laboratory for Microstructural Material Physics of Hebei Province, School of Science, Yanshan University, Qinhuangdao 066004, China*

[3]*Skolkovo Institute of Science and Technology, Bolshoy Boulevard 30, bld. 1, Moscow 121205, Russia*

[4]*Key Laboratory of Weak-Light Nonlinear Photonics, School of Physics, Nankai University, Tianjin 300071, China*

[5]*National Laboratory of Solid State Microstructures, School of Physics, and Collaborative Innovation Center of Advanced Microstructures, Nanjing University, Nanjing 210093, China*

Email: yanggc@ysu.edu.cn; xfzhou@ysu.edu.cn

The discovery of superconductivity in $CaC_6$ with a critical temperature ($T_c$) of 11.5 K reignites much interest in exploring high-temperature superconductivity in graphite intercalation compounds (GICs). Here we identify a GIC $NaC_4$, discovered by *ab initio* evolutionary structure search, as a superconductor with a computed $T_c$ of 41.2 K at 5 GPa. This value is eight times higher than that of the synthesized GIC $NaC_2$ and possesses the highest $T_c$ among available GICs. The remarkable superconductivity of GIC $NaC_4$ mainly arises from the coupling of $\pi$ electrons in graphene with the low-frequency vibrations involving both Na and C atoms. These findings suggest that Na-GICs may hold great promise as high-$T_c$ superconductors.

## I. Introduction

The search for high-temperature superconductors and the discovery of their origins are ongoing topics in condensed matter physics [1-3]. Bardeen-Cooper-Schrieffer (BCS) theory allows one to calculate properties of conventional superconductors. Compounds made of light elements usually have a high Debye temperature, which favors high-temperature superconductivity [4,5]. Thus far, pressurized hydrides, the most promising candidate, have demonstrated remarkably high critical temperatures ($T_c$s) that approach room temperature [6,7]. However, maintaining their superconductivity requires extremely high pressure of ~150 GPa or more [8,9], which presents strict requirements for scientific instruments and precludes practical applications. In this regard, the discovery of high-$T_c$ superconductors stabilized at low pressure or even at ambient pressure is the next recognized target [10,11].

Carbon is the sixth element in the periodic table and has a low atomic mass. Carbon forms the richest variety of allotropes and compounds among the light elements due to diverse hybridizations (e.g., *sp*, *sp*$^2$, and *sp*$^3$) [12-14]. An intriguing feature is that some carbon-based materials synthesized at high temperatures and high pressures can be quenchable to ambient conditions. Therefore, the investigation of superconductivity in carbon-based materials has always been in focus, and new discoveries continue to emerge [15-19]. For instance, multiple types of carbon-based superconductors have been confirmed, including B doped diamond, Q-carbon, graphite-diamond hybrid (3D carbon framework) [16-19], $YbC_6$ and $CaC_6$ (GICs, 2D carbon framework) [20,21], $Li_2C_2$ (1D carbon form) [22], and alkali metal doped $C_{60}$ (0D carbon form) [23,24].

GICs are typical layered compounds formed by inserting other atoms or molecules into interlayer spaces of graphite. The species of intercalants and their stoichiometries can modify the stacking pattern, arrangement of intercalated atoms, and the interlayer spacing. Generally, an n-stage GIC represents n successive graphene layers that are separated by the intercalant species [25]. It is evident that pressure can modify the metal concentration in GICs [26]. These features make GICs with the rich structures and intriguing properties, especially for superconductivity which is absent in graphite. Among the reported superconducting GICs, $CaC_6$ has the highest $T_c$ of 11.5 K at ambient pressure, and this increases further to 15.1 K at 7.5 GPa [20,21,27,28]. Since then, no other GICs could break this $T_c$ record.

With respect to well-studied $CaC_6$, the intercalation of sodium into graphite is also promising, partially due to the slightly smaller atomic radius and lower electronegativity, as well as different number of valence electrons. It was reported that GICs $NaC_2$ and $NaC_3$ have been synthesized at pressures ranging from 1.6 to 3.7 GPa [29]. In particular, GIC $NaC_2$ exhibits a measured $T_c$ of 5.0 K at 3.5 GPa, but its crystal structure remains unresolved. Furthermore, the stable compositions of $Na_4C$, $Na_3C_2$, $NaC$, $Na_2C_3$, and $NaC_2$ were predicted within the pressure range from 1 atm to 100 GPa. Strikingly, the structure of $P6/mmm$ $NaC_2$ is isostructural to $MgB_2$, and has a predicted $T_c$ of ~42 K at 80 GPa [30]. Moreover, the cage-based $NaC_6$ and $NaC_8$ are both superconductors with the calculated $T_c$ of 116 K and 11 K, respectively [31,32]. All of these indicate that alkali metal carbides may hold great promise as high-$T_c$ superconductors. Meanwhile, it is unknown whether there are other Na-GICs under pressure and whether they are superconducting. With these points in mind, we systematically explored various chemical compositions of potential Na-C compounds at pressures of 5 and 10 GPa, focusing on high-pressure phases that have not been previously investigated [30].

## II. Computational Methodology

Crystal structure prediction has played a major role in accelerating the discovery of new materials, especially at extreme conditions [33-38]. In this work, the variable-composition evolutionary algorithm USPEX was utilized to predict thermodynamically stable compounds in the Na-C system [33,34]. At the selected pressures of 5 and 10 GPa, we performed structure searches with an unbiased sampling of the entire range of compositions, varying the stoichiometries and their structures simultaneously. Specifically, two independent structure searches at every single pressure were performed with the number of atoms per primitive cell ranging from 6 to 20 and from 16 to 32, respectively. For each structure search, a plane-wave basis sets a cutoff of 600 eV and a grid of spacing $2\pi \times 0.06$ Å$^{-1}$ was used for Brillouin zone (BZ) sampling. The first generation was produced randomly and the fittest 40% of the population were given the probabilities to be the parent structures in the next generation-20% by heredity, 20% by lattice mutation, 10% by transmutation, and 50% were newly added random structures. The initial population consisted of 60 structures, and all other generations combined add up to ~3000 structures, thus the total number of structures is ~12000 at pressures of 5 and 10 GPa. Structure relaxations and electronic properties calculations were carried out within the framework of density functional theory [39,40] as implemented by the Vienna *ab initio* simulation package (VASP) [41]. The generalized gradient approximation (GGA) of Perdew-Burke-Ernzerhof (PBE) functional was employed for the calculation [42]. The projector augmented wave (PAW) pseudopotentials [43], with $2s^22p^63s^1$ and $2s^22p^2$ valence electrons for Na and C atoms, were used to describe the interactions between electrons and ions. A plane wave basis set with a cutoff of 1000 eV and the $k$-point meshes with a resolution better than $2\pi \times 0.022$ Å$^{-1}$ in the reciprocal space were used to ensure the total energy convergence ($10^{-6}$ eV/cell). We fully relaxed the lattice parameters and atomic coordinates until the force on each atom was less than 0.001 eV/Å.

The Quantum ESPRESSO package [44] was used to calculate lattice dynamics and electron-phonon coupling (EPC) using Optimized norm-conserving Vanderbilt pseudopotentials (ONCVPSP) [45]. The wave function cutoff energy was 150 Ry, and the charge density cutoff energy was 600 Ry. Different $k$-meshes ($q$-meshes) were chosen for the predicted compounds: $15 \times 15 \times 9$ ($5 \times 5 \times 3$) for $P2_1/m$ NaC$_2$, $12 \times 12 \times 12$ ($6 \times 6 \times 6$) for $Cmcm$ NaC$_4$ and $16 \times 16 \times 16$ ($8 \times 8 \times 8$) for $P6/mmm$ NaC$_6$. In addition, EPC calculations were also performed for GIC CaC$_6$ with norm-conserving pseudopotentials. The cutoff energy of wave functions and the $q$-mesh are adopted using 60 Ry and $6 \times 6 \times 6$, respectively. The $T_c$ value was estimated by the Allen-Dynes-modified McMillan formula [46],

$$T_c = \frac{\omega_{\log}}{1.2}\exp[-\frac{1.04(1+\lambda)}{\lambda - \mu^*(1+0.62\lambda)}],$$

where $\lambda$ is the EPC strength, $\omega_{\log}$ is the logarithmic average phonon frequency, and $\mu^*$ is the Coulomb pseudopotential parameter. The parameters $\lambda$ and $\omega_{\log}$ are defined as

$$\lambda = 2\int_0^\infty \frac{\alpha^2 F(\omega)}{\omega}d\omega,$$

and

$$\omega_{\log} = \exp[\frac{2}{\lambda}\int_0^\infty \frac{d\omega}{\omega}\alpha^2 F(\omega)\ln \omega],$$

respectively.

## III. Result and Discussion

The enthalpy of formation ($\Delta H_f$) is defined as $\Delta H_f$ (Na$_x$C$_{1-x}$) = $H$(Na$_x$C$_{1-x}$) − $xH$(Na) − $(1−x)H$(C), where $H$ represents the enthalpy of compounds or elemental solids. At a given pressure, the Na-C structures located on the convex hulls [indicated by solid lines in Fig. 1(a)] are thermodynamically stable against decomposition into other binary compounds and elemental solids. As illustrated in Fig. 1(a), NaC$_2$ with $P2_1/m$ symmetry and NaC$_6$ with $P6/mmm$ symmetry are thermodynamically stable at 5 GPa, which is partially consistent with the experimental results where the first-stage GIC NaC$_2$ and NaC$_3$ were synthesized below 4 GPa. At 10 GPa, NaC$_4$ with $Cmcm$ symmetry emerges on the convex hull, but NaC$_6$ is metastable. To provide more information for potential experimental study, the pressure-composition phase diagram was plotted in Fig. 1(b), which demonstrates the thermodynamic stability range of the predicted compounds. Specifically, $P2_1/m$ NaC$_2$ is stable in the pressure range from 4.4 to at least 10 GPa, and from 2.9 to 9.3 GPa for $P6/mmm$ NaC$_6$, whereas $Cmcm$ NaC$_4$ is stable above 8.9 GPa. Moreover, additional calculations including van der Waals (vdW) correction were carried out by using optB88-vdW [47]. As shown in Fig. S1, the inclusion of vdW interaction just slightly affect the stable pressure range on the predicted structures [48]. The pressure of formation for Na-GICs should be easily accessible within the current experimental technology, i.e., large-volume multi-anvil system or diamond anvil cell experiments. What is more important, these predicted high-pressure structures may be quenchable to ambient pressure [Figs. S2(a)-(c)] [48], giving them potential practical value.

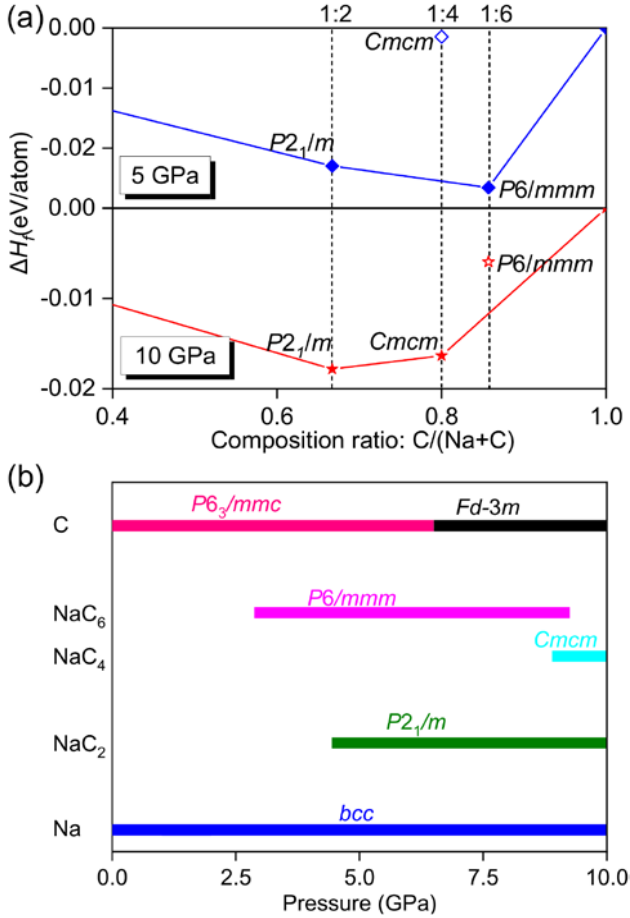

**Fig. 1.** (a) The calculated convex hulls for the Na-C systems at 5 and 10 GPa. The elemental reference structures are *bcc*-Na, graphite at 5 GPa, and diamond at 10 GPa, respectively. (b) Pressure-composition phase diagram of Na-C compounds within the pressure range from 0 to 10 GPa.

As shown in Fig. 2, the three compounds demonstrate a common structural feature: the C atoms constitute the honeycomb-like graphene, and the Na atoms are located within the voids of the graphene interlayers. As a result, all of the predicted stable phases belong to the first-stage GICs. Their lattice parameters and atomic positions at 5 GPa are listed in Table S1 [48]. $P2_1/m$ NaC$_2$ has a monoclinic structure, in which the Na atoms form double layers. Within the Na layer, the nearest Na-Na distance is ~3.28 Å. Two adjacent Na layers are interconnected by zigzag chains, in which the Na-Na distances are 3.21 and 3.32 Å, respectively [Fig. 2(a)]. *Cmcm* NaC$_4$ stabilizes into an orthorhombic structure [Fig. 2(b)] above 8.9 GPa, where the Na atoms are arranged in a zigzag-like configuration with the nearest distance ~3.12 Å [Fig. 2(b)]. Notably, metal atoms in GICs usually locate above the center of hexagonal carbon ring, but Na atoms in NaC$_4$ significantly deviate from the center. Here, we have constructed a hypothetical model of *c*-NaC$_4$ by moving the Na atoms above the center of hexagonal carbon ring. After structure relaxation, Na atoms in *c*-NaC$_4$ return to the original positions of *Cmcm* NaC$_4$. In other words, Na atoms in *Cmcm* NaC$_4$ prefer to locate above the off-center configuration. $P6/mmm$ NaC$_6$ has Na atoms arranged in a triangular form with the nearest distance of ~4.31 Å [Fig. 2(c)], similar to the B layer in $P6/mmm$ BH [49]. Moreover, the stacking sequence of graphene and Na layers in $P6/mmm$ NaC$_6$ is A$\alpha$A$\alpha$, which differs from that in $R$-$3m$ CaC$_6$, where it is A$\alpha$A$\beta$A$\gamma$A [50]. The C-C bonding lengths in the Na-GICs are slightly larger than those in pristine graphene (Table S2) [48,51], which can be attributed to the charge transfer from Na to C, leading to the electronic occupation of C-C antibonding orbital, weakening the covalent C-C bonds [Figs. S3(a)-(c)] [48].

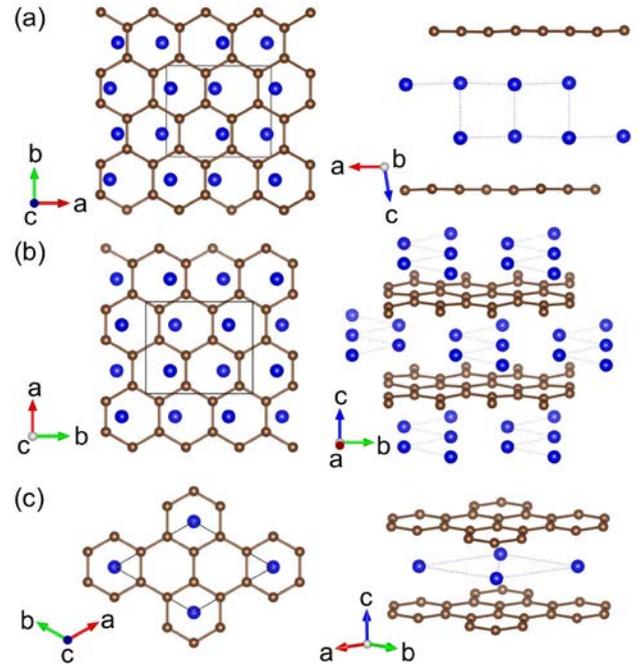

**Fig. 2.** Crystal structures of Na-GICs at 5 GPa. (a) $P2_1/m$ NaC$_2$, (b) *Cmcm* NaC$_4$, and (c) $P6/mmm$ NaC$_6$. For these structures, the blue and brown spheres represent Na and C atoms, respectively.

Inspired by the unique structures of the three predicted Na-GICs, we proceeded to explore their electronic properties through calculating projected electronic band structures and density of states (DOS). The three phases demonstrate intrinsic metallicity with several bands crossing the Fermi level [$E_F$, see Fig. 3(a) and Fig. S4(a)-(d)] [48]. The normalized electronic DOS at $E_F$ are ~0.02 states/(eV·Å$^3$) for NaC$_2$ and NaC$_6$ at 5 GPa, whereas ~0.024 states/(eV·Å$^3$) for NaC$_4$. Among three Na-GICs, the Fermi level of NaC$_4$ is closest to the Van Hove singularities [Fig. 3(b), Figs. S4(b) and S4(d)], implying that it may have better superconductivity [48]. Subsequently, we focused on analyzing the electronic properties of *Cmcm* NaC$_4$. At 5 GPa, its metallicity mainly arises from the C $p_z$ orbital electrons [Fig. 3(b)], which form a system of delocalized $\pi$ bonds. The delocalized $\pi$ electrons in the honeycomb-like graphene play a critical role in metallicity. One can notice steep bands along the S-R,

R-Z, and Z-T directions and flat bands at the high symmetry points T and Γ near $E_F$ [Fig. 3(b)], signifying the high electron velocity and large DOS.

The topology of the Fermi surface is helpful in understanding the behavior of electrons at $E_F$. For *Cmcm* NaC$_4$, there are three bands crossing $E_F$ [Fig. 3(c) and Fig. S5] [48]. Here, we explore band1 and band2, which make the dominant contribution to the Fermi surface. The Fermi surface of band1 consists of eight sheets, whereas band2 is composed of one "8"-type and two U-type sheets. Besides the minor contribution of Na *p* states to the Fermi surface from band2, the two Fermi surfaces are mainly derived from the C $p_z$ states. More interestingly, the two Fermi surfaces are nested along the body diagonal and Γ-Z/S/Y direction of Brillouin zone (BZ). It is known that nesting can lead to a superconductivity or instability. It could be in favor of EPC since NaC$_4$ is mechanically and dynamically stable under pressure [Fig. 3(c)].

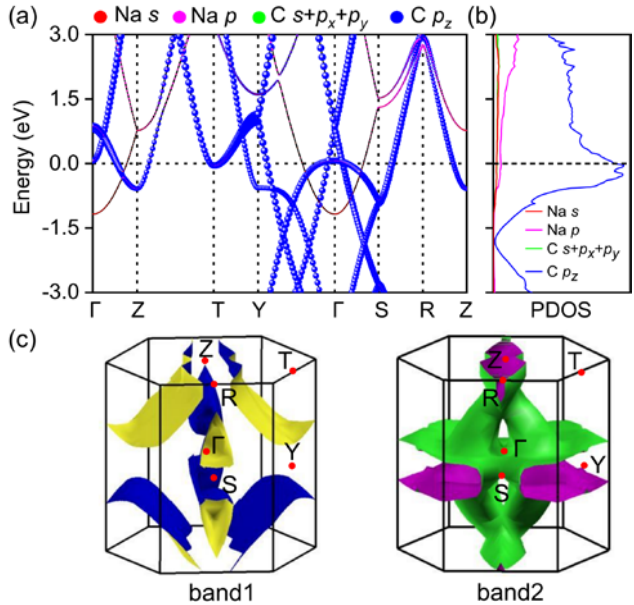

**Fig. 3.** (a) The orbital-resolved band structures of *Cmcm* NaC$_4$ at 5 GPa. (b) Projected density of states (PDOS) (the dashed line indicates $E_F$). (c) The Fermi surfaces associated with two bands crossing $E_F$.

To establish the reliability of our computation method, we first employed the Allen-Dynes modified McMillan equation to estimate the superconductivity of GIC CaC$_6$. The calculated values of $\lambda$, $\omega_{\log}$, and $T_c$ are 0.75, 386.9 K and 11.3 K, respectively, with $\mu^* = 0.14$, which is in good agreement with both theoretical and experimental results [20,21,52,53]. For Na-GICs, superconductivity of $P2_1/m$ NaC$_2$ at 3.5 GPa was also calculated with $\mu^* = 0.1$ [Fig. S(6)] [48]. The computed $T_c$ of 5.4 K is in excellent agreement with the measured value of 5.0 K in NaC$_2$ at 3.5 GPa [29]. By contrast, the calculated EPC parameter $\lambda$ of *Cmcm* NaC$_4$ is 1.01 at 5 GPa, comparable to MgB$_2$ (1.0 at 0 GPa) [54]. The phonon dispersion curves with $\lambda$ weights show strong EPC in the range of 0-479 and 1200-1448 cm$^{-1}$ in the whole BZ [Fig. 4(a)], especially in the range of 0-479 cm$^{-1}$ along the Γ-Z, Γ-Y, and Γ-S directions. This result mostly related to phonon softening consistent with the Fermi surface nesting, supported by the distinct sharp peaks of the Fermi surface nesting function $\xi_q$ [Fig. 4(d)]. By comparing the Eliashberg spectral function $\alpha^2F(\omega)$ and PHDOS, we found that low frequency phonons (below 479 cm$^{-1}$), associated with Na and C coupling vibrations, contribute 70% to $\lambda$, while high-frequency phonons (479-1448 cm$^{-1}$) contribute 30% [Fig. 4(b)-4(c)]. The latter is related to vibrations of strong covalent C-C bonds. As a result, superconductivity of NaC$_4$ predominantly originates from the coupling of C $p_z$ electrons with the low-frequency phonons. The estimated $T_c$ is 41.2 K at 5 GPa with a value of $\mu^* = 0.1$, making it the highest among the reported GICs. The pressure-dependent superconductivity of NaC$_4$ at the pressures of 0, 5, and 10 GPa is also investigated [Fig. 4(e)]. At zero pressure, the estimated $T_c$ of *Cmcm* NaC$_4$ is 36.6 K. With increasing pressure, $T_c$ rises first (41.2 K at 5 GPa) and then falls (38.3 K at 10 GPa), which can be explained by the variations of $\omega_{\log}$ and $\lambda$. From 0 to 5 GPa, $\lambda$ remains unchanged for *Cmcm* NaC$_4$, while $\omega_{\log}$ is significantly enhanced due to phonon stiffening, leading to a higher $T_c$ value. As the pressure increases from 5 to 10 GPa, both $\omega_{\log}$ and $\lambda$ gradually decrease, leading to the decline of $T_c$.

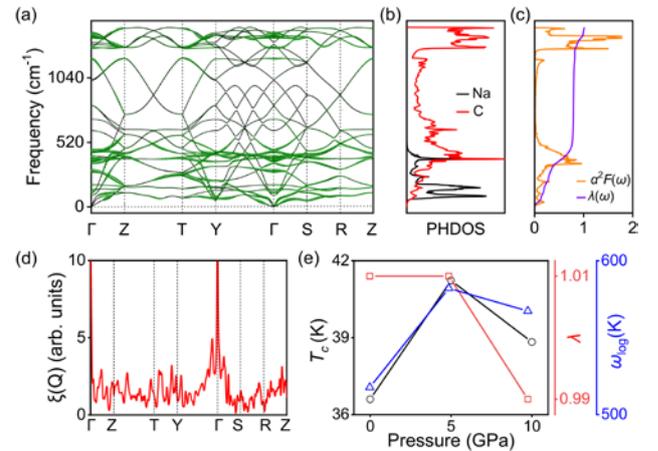

**Fig. 4.** (a) Phonon dispersion curves of *Cmcm* NaC$_4$ at 5 GPa (the magnitude of $\lambda_{q,\nu}$ indicated by the thickness of the green curves). (b) Projected phonon DOS(PHDOS). (c) Eliashberg spectral function $\alpha^2F(\omega)$ (orange line), frequency-dependent EPC parameter $\lambda(\omega)$ (purple line). (d) The Fermi surface nesting function $\xi_q$ along some *q* trajectories. (e) Pressure-depended $T_c$, $\omega_{\log}$, and $\lambda$ of *Cmcm* NaC$_4$.

On the other hand, the superconductivity of NaC$_2$ and NaC$_6$ is also analyzed and compared to that of NaC$_4$ at ambient pressure [Fig. S2, Fig. S7] [48]. The resulting $\lambda$ values are 0.58 and 0.62 for NaC$_2$ and NaC$_6$, respectively, which are significantly lower than that of NaC$_4$ (1.01). The corresponding $T_c$ values are 13.34 K, 24.11 K, and 36.6 K for NaC$_2$, NaC$_6$ and NaC$_4$, respectively. To explore the superior superconductivity of NaC$_4$ than NaC$_2$ and NaC$_6$, the EPC strength $\lambda$ can be approximately divided into two parts: the partial $\lambda_\mathrm{I}$ is contributed by the coupled vibrations between Na and C, whereas partial $\lambda_\mathrm{II}$ is contributed by the vibrations of C atoms (Fig. S2 and Table S3) [48]. According to this definition, the ratio of $\lambda_\mathrm{I}$ ($\lambda_\mathrm{II}$) to the EPC strength $\lambda$ are 29.3% (70.7%), 61.4% (38.6%), and 21% (79%) for NaC$_2$, NaC$_4$, and NaC$_6$ at ambient pressure. It is evident that $\lambda_\mathrm{II}$ dominates the superconductivity for NaC$_2$ and NaC$_6$ while $\lambda_\mathrm{I}$ plays a crucial role in the superconductivity on NaC$_4$. Therefore, even though the carbon sublattice is similar in the three Na-GICs, the difference in the concentration and configuration of Na substantially modulates the EPC (Table S4) [48].

## IV. Conclusions

In summary, three Na-GICs, NaC$_2$, NaC$_4$, and NaC$_6$, were predicted from *ab initio* evolutionary structure search. NaC$_2$ comprises double Na layers, NaC$_4$ has zigzag Na chains, and Na layers in NaC$_6$ feature a triangular lattice configuration. Among them, NaC$_4$ demonstrates a $T_\mathrm{c}$ of 41.2 K at 5.0 GPa, which sets a new record for reported GICs. The superconductivity of NaC$_4$ mainly originates from the coupling of C $p_z$ electrons with the low-frequency phonons (the coupled vibrations between Na and C), which is distinct from that of NaC$_6$ (the coupling of C $p_z$ electrons with C-derived vibrations). Strikingly, the predicted $T_\mathrm{c}$ value of NaC$_2$ is very close to the measured one, suggesting that the long-lasting puzzle surrounding the structure of experimental NaC$_2$ may have been resolved. Our findings shed light on the exploration of high-$T_\mathrm{c}$ superconductors in similar GICs.

## Acknowledgments

This work was supported by the National Key R&D Program of China (Grant No. 2022YFA1402300), National Natural Science Foundation of China (Grants No. 52025026, No. 52090020, No. 12174200 and No. 92263101). A.R.O. acknowledges funding from the Russian Science Foundation (Grant No. 19-72-30043).

**References**

[1] J. G. Bednorz and K. A. Müller, Z. Phys. B **64**, 189 (1986).

[2] J. Nagamatsu, N. Nakagawa, T. Muranaka, Y. Zenitani, and J. Akimitsu, Nature **410**, 63 (2001).

[3] L. Ma, K. Wang, Y. Xie, X. Yang, Y. Wang, M. Zhou, H. Liu, X. Yu, Y. Zhao, H. Wang, G. Liu, and Y. Ma, Phy. Rev. Lett. **128**, 167001 (2022).

[4] N. W. Ashcroft, Phys. Rev. Lett. **21**, 1748 (1968).

[5] N. W. Ashcroft, Phys. Rev. Lett. **92**, 187002 (2004).

[6] P. Kong, V. S. Minkov, M. A. Kuzovnikov, A. P. Drozdov, S. P. Besedin, S. Mozaffari, L. Balicas, F. F. Balakirev, V. B. Prakapenka, S. Chariton, D. A. Knyazev, E. Greenberg, M. I. Eremets, Nat. Commun. **12**, 5075 (2021).

[7] M. Somayazulu, M. Ahart, A. K. Mishra, Z. M. Geballe, M. Baldini, Y. Meng, V. V. Struzhkin, and R. J. Hemley, Phys. Rev. Lett. **122**, 027001 (2019).

[8] Z. Zhang, T. Cui, M. J. Hutcheon, A. M. Shipley, H. Song, M. Du, V. Z. Kresin, D. Duan, C. J. Pickard, and Y. Yao, Phys. Rev. Lett. **128**, 047001 (2022).

[9] X. Zhang, Y. Zhao, and G. Yang, WIREs Comput. Mol. Sci. **12**, e1582 (2022).

[10] W. Chen, D. V. Semenok, X. Huang, H. Shu, X. Li, D. Duan, T. Cui, and A. R. Oganov, Phys. Rev. Lett. **127**, 117001 (2021).

[11] X. Liang, A. Bergara, X. Wei, X. Song, L. Wang, R. Sun, H. Liu, R. J. Hemley, L. Wang, G. Gao, and Y. Tian, Phys. Rev. B **104**, 134501 (2021).

[12] F. Lavini, M. Rejhon, and E. Riedo, Nat. Rev. Mater. **7**, 814 (2022).

[13] E. D. Miller, D. C. Nesting, and J. V. Badding, Chem. Mater **9**, 18 (1997).

[14] B. Sundqvist, Physics Reports **909**, 1 (2021).

[15] H. Zhou, T. Xie, T. Taniguchi, K. Watanabe, and A. F. Young, Nature **598**, 434 (2021).

[16] E. A. Ekimov, V. A. Sidorov, E. D. Bauer, N. N. Mel'nik, N. J. Curro, J. D. Thompson, and S. M. Stishov, Nature **428**, 542 (2004).

[17] A. Bhaumik, R. Sachan, and J. Narayan, ACS Nano **11**, 5351 (2017).

[18] A. Bhaumik, R. Sachan, S. Gupta, and J. Narayan, ACS Nano **11**, 11915 (2017).

[19] Y. Ge, K. Luo, Y. Liu, G. Yang, W. Hu, B. Li, G. Gao, X.-F. Zhou, B. Xu, Z. Zhao, and Y. Tian, Mater. Today Phys. **23**,100630 (2022).

[20] T. E. Weller, M. Ellerby, S. S. Saxena, R. P. Smith, and N.T. Skipper, Nat. Phys. **1**, 39 (2005).

[21] N. Emery, C. Hérold, M. d'Astuto, V. Garcia, Ch. Bellin, J. F. Marêché, P. Lagrange, and G. Loupias, Phys. Rev. Lett. **95**, 087003 (2005).

[22] M. Gao, X. Kong, Z. Lu, and T. Xiang, Acta Phys. Sin. **64**, 214701 (2015).

[23] A. Y. Ganin, Y. Takabayashi, Y. Z. Khimyak, S. Margadonna, A. Tamai, M. J. Rosseinsky, and K. Prassides, Nat. Mater. **7**, 367 (2008).

[24] A. F. Hebard, M.J. Rosseinsky, R.C. Haddon, D.W. Murphy, S.H. Glarum, Palstra, Thomas, A.P. Ramirez, and A.R. Kortan, Nature **350**, 600 (1991).

[25] M. S. Dresselhaus, G. Dresselhaus, Adv. Phys. **51**, 1 (2002).

[26] I. T. Belash, O. V. Zharikov, and A. V. Palnichenko, Synth. Met. **34**, 47 (1990).

[27] A. Gauzzi, S. Takashima, N. Takeshita, C. Terakura, H. Takagi, N. Emery, C. Hérold, P. Lagrange, and G. Loupias, Phys. Rev. Lett. **98**, 067002 (2007).

[28] G. Csányi, P. B. Littlewood, A. H. Nevidomskyy, C. J. Pickard and B. D. Simons, Nat. Phys. **1**, 42 (2005).

[29] I. T. Belash, A. D. Bronnikov, O. V. Zharikov, and A.V. Palnichenko, Solid State Commun. **64**, 1445 (1987).

[30] Q. Yang, K. Zhao, H. Liu, and S. Zhang, J. Phys. Chem. Lett. **12**, 5850 (2021).

[31] S. Lu, H. Liu, I. I. Naumov, S. Meng, Y. Li, J. S. Tse, B. Yang, and R. J. Hemley, Phys. Rev. B **93**, 104509 (2016).

[32] J.-Y. You, B. Gu, and G. Su, Phys. Rev. B **101**, 184521 (2020).

[33] A. R. Oganov and C. W. Glass, J. Chem. Phys. **124**, 244704 (2006).

[34] A. O. Lyakhov, A. R. Oganov, H. T. Stokes, and Q. Zhu, Comput. Phys. Commun. **184**, 1172 (2013)

[35] Y. Wang, J. Lv, L. Zhu, and Y. Ma, Comput. Phys. Commun. **183**, 2063 (2012).

[36] C. J. Pickard, and R. J. Needs, J. Phys. Condens. Mat. **23**, 053201 (2011).

[37] S. Goedecker, J. Chem. Phys. **120**, 9911 (2004).

[38] M. Amsler, and S. Goedecker, J. Chem. Phys. **133**, 224104 (2010).

[39] W. Kohn and L. J. Sham, Phys. Rev. **140**, A1133 (1965).

[40] P. Hohenberg and W. Kohn, Phys. Rev. **136**, B864 (1964).

[41] G. Kresse and J. Furthmüller, Phys. Rev. B **54**, 11169 (1996).

[42] J. P. Perdew, K. Burke, and M. Ernzerhof, Phys. Rev. Lett. **77**, 3865 (1996).

[43] P. E. Blöchl, Phys. Rev. B **50**, 17953 (1994).

[44] P. Giannozzi, S. Baroni, N. Bonini, M. Calandra, R. Car, C. Cavazzoni, D. Ceresoli, G. L. Chiarotti, M. Cococcioni, and I. Dabo, J. Phys. Condens. Mat. **21**, 395502 (2009).

[45] D. R. Hamann, Phys. Rev. B **88**, 085117 (2013).

[46] P. B. Allen, and R. C. Dynes, Phys. Rev. B **12**, 905 (1975).

[47] J. Klimes, D. R. Bowler, and A. Michaelides, J. Phys.: Condens. Matter **22**, 022201 (2010).

[48] See Supplemental Material for additional electronic properties.

[49] C. H. Hu, A. R. Oganov, Q. Zhu, G. R. Qian, G. Frapper, A. O. Lyakhov, and H. Y. Zhou, Phys. Rev. Lett. **110**, 165504 (2013).

[50] N. Emery, C. Hérold, and P. Lagrange, J. Solid State Chem. **178**, 2947 (2005).

[51] M. F. Budyka, T. S. Zyubina, A. G. Ryabenko, S. H. Lin, and A. M. Mebel, Chem. Phys. Lett. **407**, 266 (2005).

[52] M. Calandra and F. Mauri, Phys. Rev. Lett. **95**, 237002 (2005).

[53] W. Chen, J. Appl. Phys. **114**, 173906 (2013).

[54] J. M. An and W. E. Pickett, Phys. Rev. Lett. **86**, 4366 (2001).

Supplementary Materials

# Superconductivity in graphite intercalation compounds with sodium

Chun-Mei Hao[1,2], Xing Li[2], Artem R. Oganov[3], Jingyu Hou[4], Shicong Ding[2], Yanfeng Ge[2], Lin Wang[1,2], Xiao Dong[4], Hui-Tian Wang[5], Guochun Yang[2,*], Xiang-Feng Zhou[1,2,4*], and Yongjun Tian[1]

[1]*Center for High Pressure Science, State Key Laboratory of Metastable Materials Science and Technology, Yanshan University, Qinhuangdao, 066004, China*

[2]*Key Laboratory for Microstructural Material Physics of Hebei Province, School of Science, Yanshan University, Qinhuangdao 066004, China*

[3]*Skolkovo Institute of Science and Technology, Bolshoy Boulevard 30, bld. 1, Moscow 121205, Russia*

[4]*Key Laboratory of Weak-Light Nonlinear Photonics, School of Physics, Nankai University, Tianjin 300071, China*

[5]*National Laboratory of Solid State Microstructures, School of Physics, and Collaborative Innovation Center of Advanced Microstructures, Nanjing University, Nanjing 210093, China*

Email: yanggc@ysu.edu.cn; xfzhou@ysu.edu.cn

| **Index** | **Page** |

# Supplementary Figures

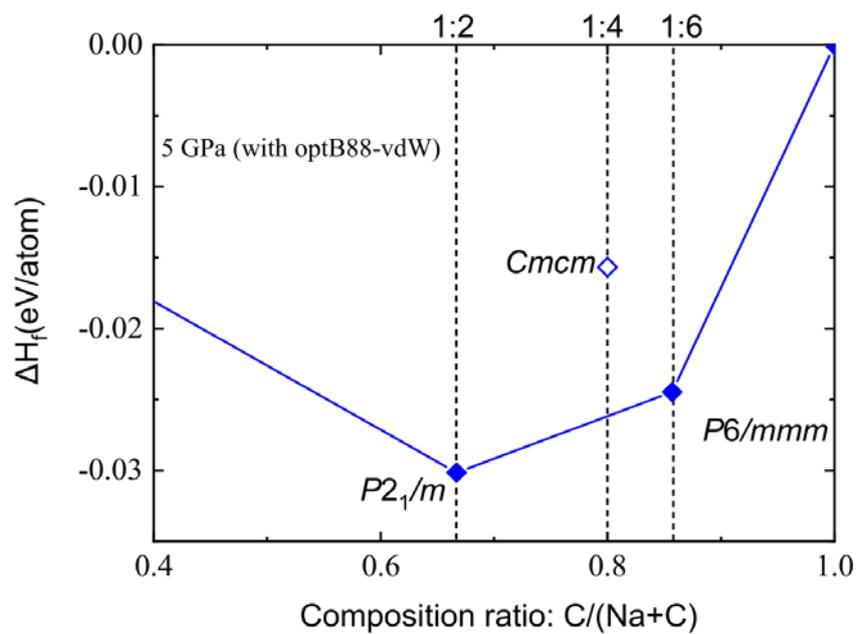

**Fig. S1.** The calculated convex hull with the inclusion of van der Waals (vdW) interactions for $P2_1/m$ NaC$_2$, $Cmcm$ NaC$_4$, and $P6/mmm$ NaC$_6$ relative to bcc Na and graphite at 5 GPa.

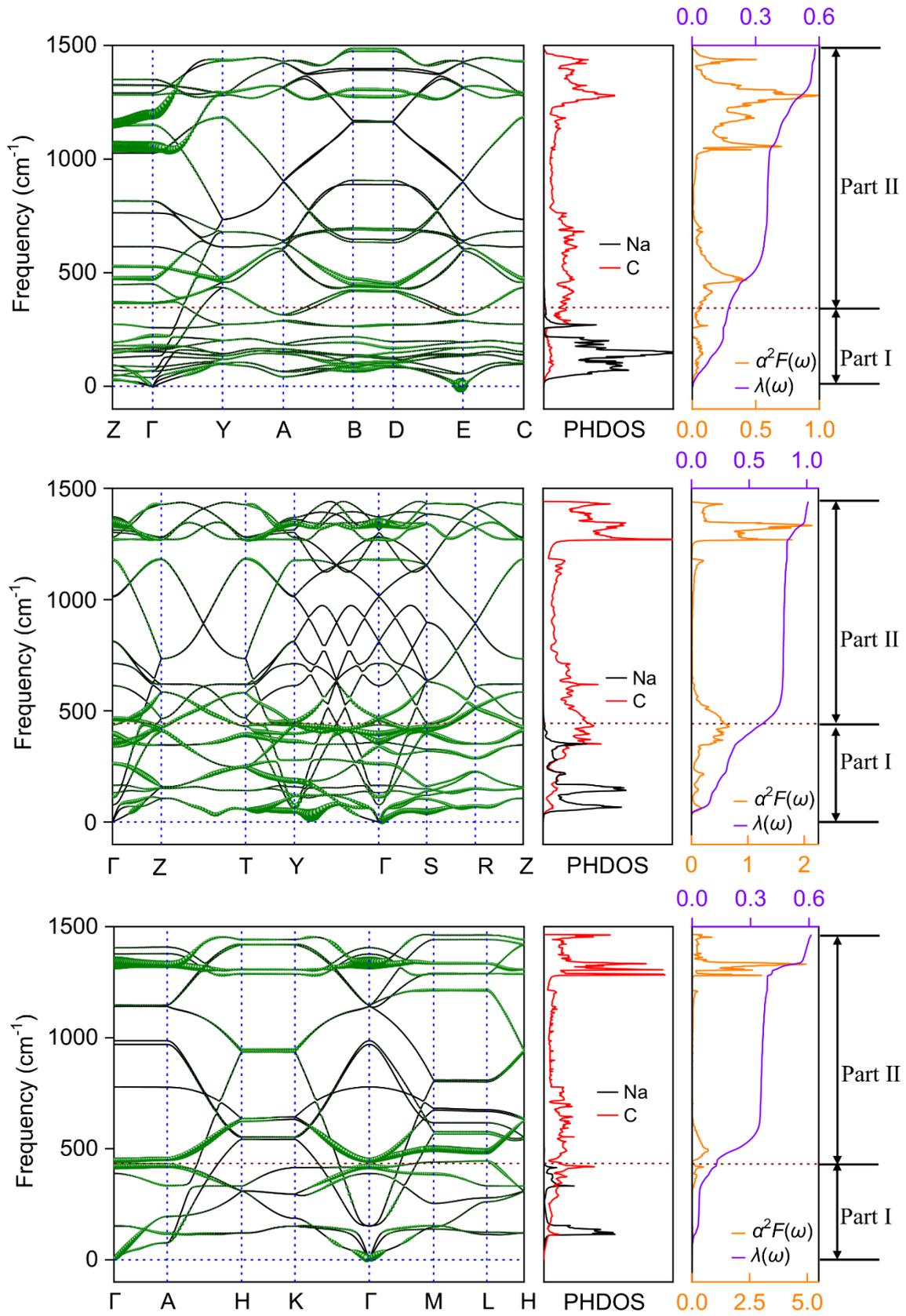

**Fig. S2.** Phonon dispersion curves (the magnitude of $\lambda_{q,v}$ indicated by the thickness of the green curves), projected phonon DOS, Eliashberg spectral function $\alpha^2F(\omega)$, and EPC parameter $\lambda$ of (a) $P2_1/m$ NaC$_2$, (b) $Cmcm$ NaC$_4$, and (c) $P6/mmm$ NaC$_6$ at 0 GPa. The PHDOS can be generally divided into two parts (labeled by the brown dashed line): One is the coupled vibrations between Na and C atoms, and the other is the vibrations of C atoms. Accordingly, Part I and Part II represent the frequency range contributed by the coupling vibrations and the vibrations of C atoms, respectively.

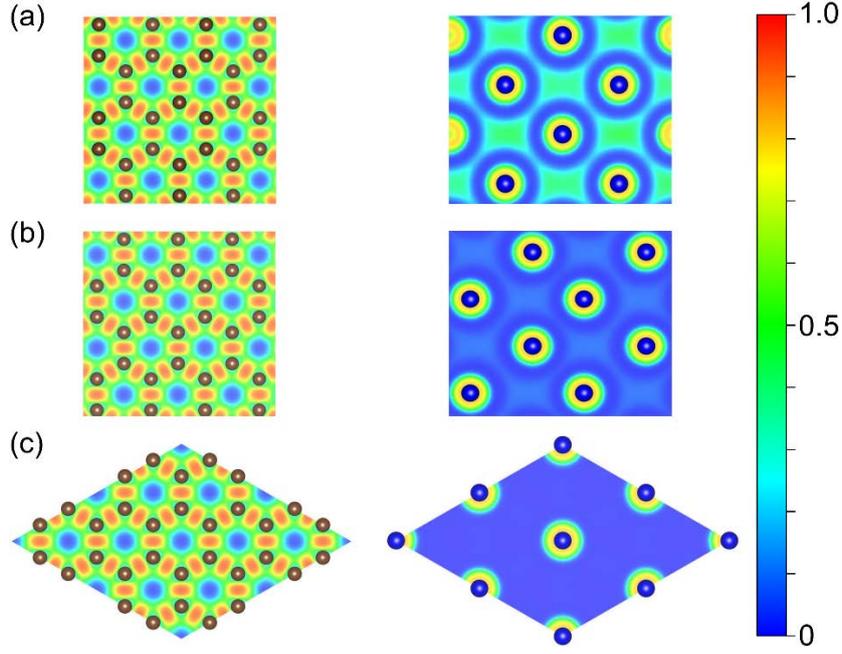

**Fig. S3.** Electron localization function (ELF) for (a) $P2_1/m$ NaC$_2$, (b) $Cmcm$ NaC$_4$, and (c) $P6/mmm$ NaC$_6$ at 5 GPa. The blue and brown spheres represent Na and C atoms, respectively. The left and right sides represent the ELF of graphene layer and Na layer, respectively.

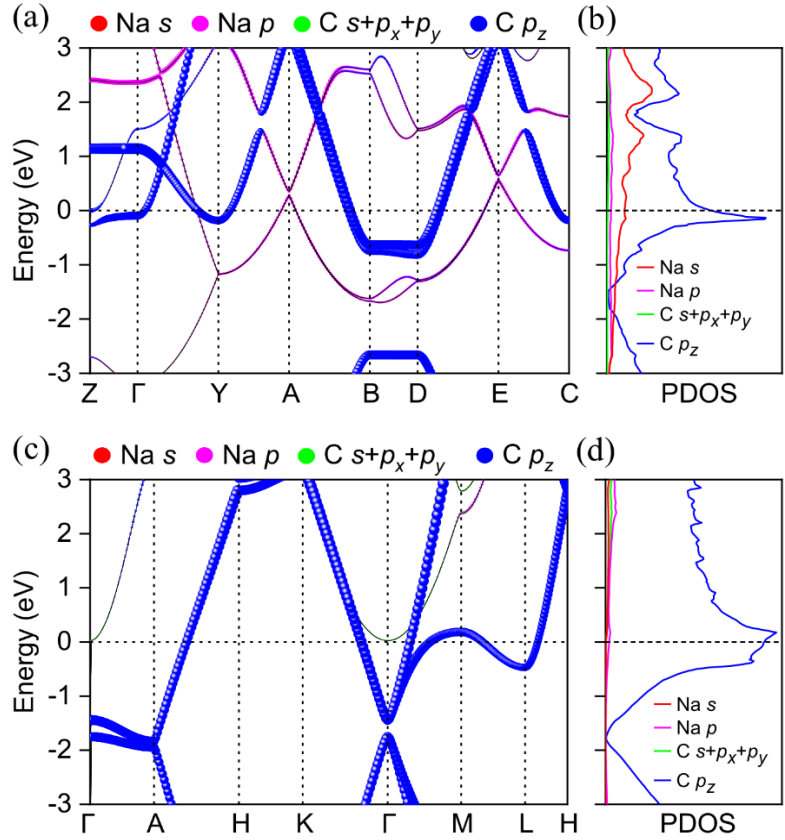

**Fig. S4.** (a), (b) The orbital-resolved band structures and projected density of states (PDOS) for $P2_1/m$ NaC$_2$. (c), (d) for $P6/mmm$ NaC$_6$ at 5 GPa.

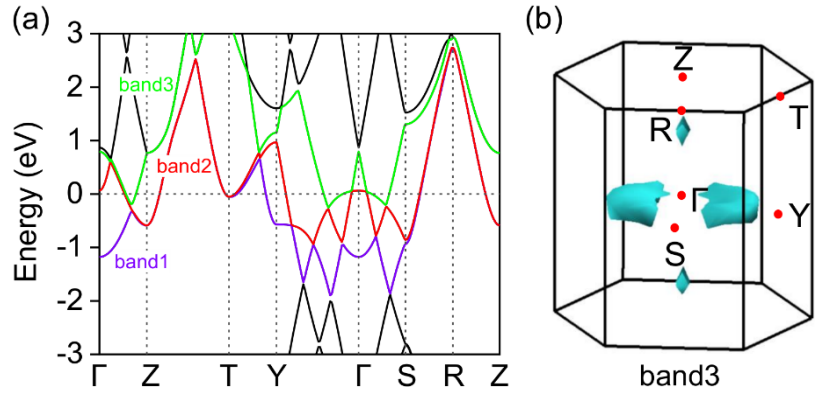

**Fig. S5.** (a) Band structure of *Cmcm* NaC$_4$ at 5 GPa. Three bands cross the Fermi level $E_F$, marked as band1 (red line), band2 (purple line) and band3 (green), respectively. (b) The Fermi surface is related to band3.

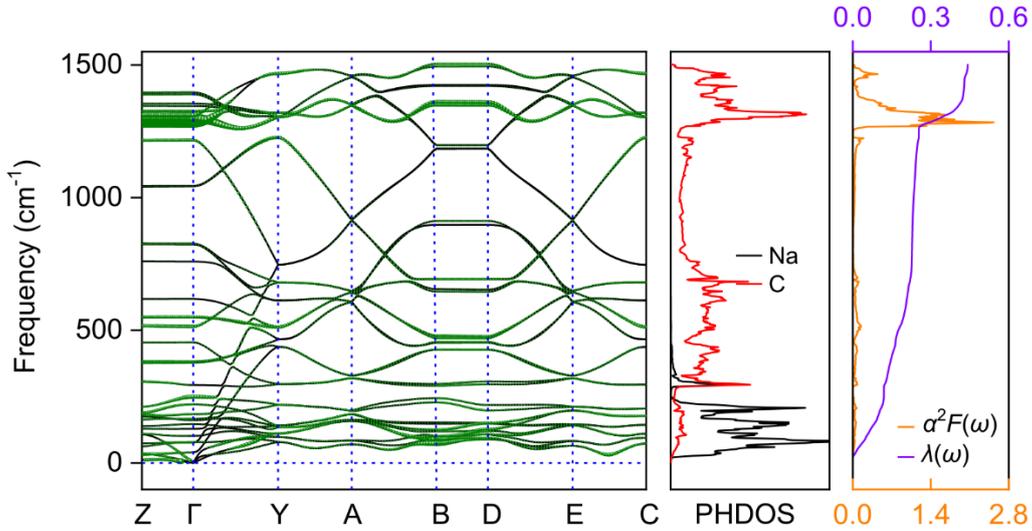

**Fig. S6.** Phonon dispersion curves (the magnitude of $\lambda_{q,v}$ indicated by the thickness of the green curves), projected phonon DOS, Eliashberg spectral function $\alpha^2F(\omega)$, and EPC parameter $\lambda$ of *P2$_1$/m* NaC$_2$ at 3.5 GPa.

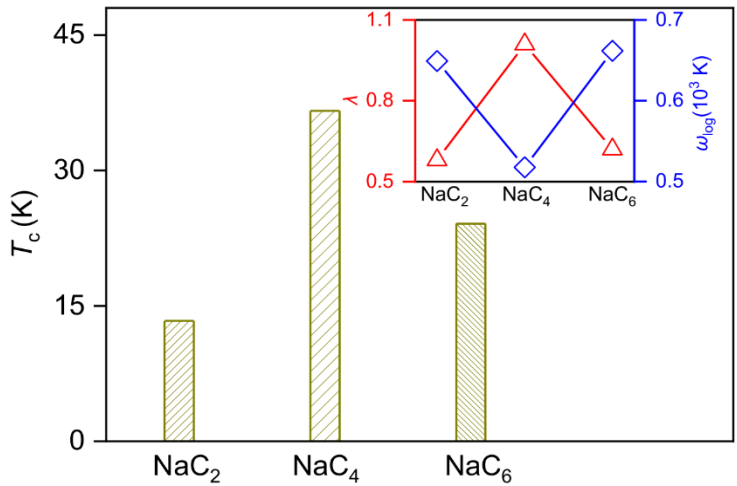

**Fig. S7.** The superconductivity of *P2$_1$/m* NaC$_2$, *Cmcm* NaC$_4$, and *P6/mmm* NaC$_6$ at 0 GPa.

S6

# Supplementary Tables

Table S1. Lattice parameters and atomic positions of stable Na-C compounds at 5 GPa. It should be mentioned that "setting c̄ba" is performed to orient the graphene layers stacking in NaC$_4$ along the *c*-axis.

| Space group (No.) | Lattice parameters (Å) | Atomic fractional coordinates |
|---|---|---|
| *P*2$_1$/*m*-NaC$_2$ (11) | a = 4.97, b = 4.30, c = 7.36, α = γ = 90°, β = 98.68° | Na1 2*e* (0.4638, 0.7500, 0.3358)<br>Na2 2*e* (0.9597, 0.2500, 0.3262)<br>C1 4*f* (0.3742, 0.9171, 0.9961)<br>C2 4*f* (0.8751, 0.9169, 0.0006) |
| *Cmcm*-NaC$_4$ (63, setting c̄ba) | a = 4.32, b = 4.99, c = 9.12, α = β = γ = 90° | Na1 4*c* (-0.2500 0.2740 -0.5000)<br>C1 16*h* (0.0830 -0.1251 -0.24520) |
| *P*6/*mmm*-NaC$_6$ (191) | a = b = 4.31, c = 4.34, α = β = 90°, γ = 120° | Na1 1*b* (0.0000 0.0000 0.5000)<br>C1 6*j* (0.0000 0.3337 0.0000) |

Table S2. The average bond lengths of C-C in Na-C compounds at 5 GPa compared to experimental data of graphene (1.42 Å).

| Space group | *P*2$_1$/*m*-NaC$_2$ | *Cmcm*-NaC$_4$ | *P*6/*mmm*-NaC$_6$ |
|---|---|---|---|
| Bond length (Å) | 1.44 | 1.44 | 1.43 |

Table S3. The EPC strength $\lambda$, the partial $\lambda_I$ contributed by the coupled vibrations between Na and C atoms, and partial $\lambda_{II}$ contributed by the vibrations of C atoms at ambient pressure; note that $\lambda = \lambda_I + \lambda_{II}$.

| Phase | $\lambda_I$ | $\lambda_{II}$ | $\lambda$ |
|---|---|---|---|
| NaC$_2$ | 0.17 | 0.41 | 0.58 |
| NaC$_4$ | 0.62 | 0.39 | 1.01 |
| NaC$_6$ | 0.13 | 0.49 | 0.62 |

Table S4. Calculated values of densities of states at the Fermi level of N ($E_f$), EPC strength $\lambda$, logarithmic average phonon frequencies $\omega_{\log}$, and superconducting critical temperatures $T_c$ with a Coulomb potential of 0.1.

| Phases | Pressure (GPa) | $N(E_f)$ (states/(eV·Å$^3$)) | $\omega_{\log}$ (K) | $\lambda$ | $T_c$ (K) |
|---|---|---|---|---|---|
| NaC$_2$ | 0 | 0.022 | 649.02 | 0.58 | 13.34 |
| | 5 | 0.021 | 768.09 | 0.43 | 4.50 |
| | 10 | 0.022 | 705.45 | 0.44 | 5.00 |
| NaC$_4$ | 0 | 0.024 | 517.70 | 1.01 | 36.60 |
| | 5 | 0.024 | 582.08 | 1.01 | 41.23 |
| | 10 | 0.022 | 567.32 | 0.99 | 38.33 |
| NaC$_6$ | 0 | 0.023 | 999.17 | 0.62 | 24.11 |
| | 5 | 0.021 | 1199.51 | 0.44 | 8.28 |
| | 10 | 0.020 | 1581.00 | 0.28 | 0.41 |

<span style="color:red">DFT parameters of EPC calculation for NaC$_4$</span>

```
&control
    calculation = 'scf',
    wf_collect=.true.,
    restart_mode='from_scratch',
    prefix='NaC',
    pseudo_dir = './',
    outdir='./temp',
    tstress = .true.,
    tprnfor = .true.,
 /
&system
  vdw_corr='DFT-D',
  ibrav=0,
  nat=10,
  ntyp=2,
  a = 1.00
  ecutwfc=150,
  ecutrho=600,
  occupations='smearing',
  smearing='methfessel-paxton',
  degauss=0.05,
 /
 &electrons
    mixing_beta = 0.5,
    conv_thr =   1.0d-14,
 /
ATOMIC_SPECIES
   Na 22.990      Na_ONCV_PBE-1.0.upf
   C  12.011      C_ONCV_PBE-1.0.upf
CELL_PARAMETERS {alat}
    4.470169773   -2.484516599    0.000000000
    4.470169773    2.484516599    0.000000000
    0.000000000    0.000000000    4.312954583
ATOMIC_POSITIONS (crystal)
Na            0.2019443604         0.7980556396         0.7500000000
Na            0.7980556396         0.2019443604         0.2500000000
C             0.3693055043         0.1193772239         0.4167676936
C             0.8806227761         0.6306944957         0.0832323064
C             0.1193772239         0.3693055043         0.9167676936
C             0.3693055043         0.1193772239         0.0832323064
C             0.1193772239         0.3693055043         0.5832323064
C             0.8806227761         0.6306944957         0.4167676936
C             0.6306944957         0.8806227761         0.9167676936
C             0.6306944957         0.8806227761         0.5832323064
K_POINTS {automatic}
12 12 12 0 0 0
```